\documentclass[conference]{IEEEtran}
\usepackage{cite}
\usepackage[cmex10]{amsmath}
\usepackage{array}
\usepackage{amsfonts}
\usepackage{mathrsfs}
\usepackage{arydshln}
\usepackage{slashbox}
\usepackage{graphicx}
\usepackage{float}
\usepackage{subfigure}
\usepackage{amssymb}
\usepackage{amsmath}
\usepackage{geometry}
\usepackage[colorlinks,linkcolor=black,urlcolor=black,anchorcolor=black,citecolor=black,hyperfootnotes=true]{hyperref}
\usepackage{multirow}
\usepackage{multicol}
\usepackage{cite}
\usepackage[subfigure]{graphfig}
\usepackage{xcolor}
\usepackage{setspace}
\newcommand{\mv}[1]{\mbox{\boldmath{$ #1 $}}}
\usepackage[linesnumbered,ruled]{algorithm2e}

\newcommand{\qed}{\nobreak \ifvmode \relax \else
      \ifdim\lastskip<1.5em \hskip-\lastskip
      \hskip1.5em plus0em minus0.5em \fi \nobreak
      \vrule height0.75em width0.5em depth0.25em\fi}

\begin{document}
\title{{Radio Map Based Path Planning for Cellular-Connected UAV}
\author{\IEEEauthorblockN{Shuowen~Zhang and Rui~Zhang}
		\IEEEauthorblockA{ECE Department, National University of Singapore. Email: \{elezhsh,elezhang\}@nus.edu.sg}}}
\maketitle
\begin{abstract}
In this paper, we study the path planning for a cellular-connected unmanned aerial vehicle (UAV) to minimize its flying distance from given initial to final locations, while ensuring a target link quality in terms of the large-scale channel gain with each of its associated ground base stations (GBSs) during the flight. To this end, we propose the use of \emph{radio map} that provides the information on the large-scale channel gains between each GBS and uniformly sampled locations on a three-dimensional (3D) grid over the region of interest, which are assumed to be time-invariant due to the generally static and large-size obstacles therein (e.g., buildings). Based on the given radio maps of the GBSs, we first obtain the optimal UAV path by solving an equivalent shortest path problem (SPP) in graph theory. To reduce the computation complexity of the optimal solution, we further propose a grid quantization method whereby the grid points in each GBS's radio map are more coarsely sampled by exploiting the spatial channel correlation over neighboring grids. Then, we solve the approximate SPP over the reduced-size radio map (graph) more efficiently. Numerical results show that the proposed solutions can effectively minimize the flying distance of the UAV subject to its communication quality constraint. Moreover, a flexible trade-off between performance and complexity can be achieved \hbox{by adjusting the quantization ratio for the radio map.}
\end{abstract}

\section{Introduction}
The applications of unmanned aerial vehicles (UAVs) have become increasingly popular and diversified, ranging from cargo delivery to aerial video streaming and virtual/augmented reality \cite{access}. To enable the safe fly of UAVs as well as to support timely exchange of mission data between them and their ground users, it is crucial to establish high-quality air-ground communications. To this end, a promising technology is \emph{cellular-connected UAV}, by leveraging the ground base stations (GBSs) in the cellular network to serve the UAVs as new users in the sky \cite{cellularUAV_arXiv}. 

\begin{figure}[t]
	\centering
	\includegraphics[width=6.5cm]{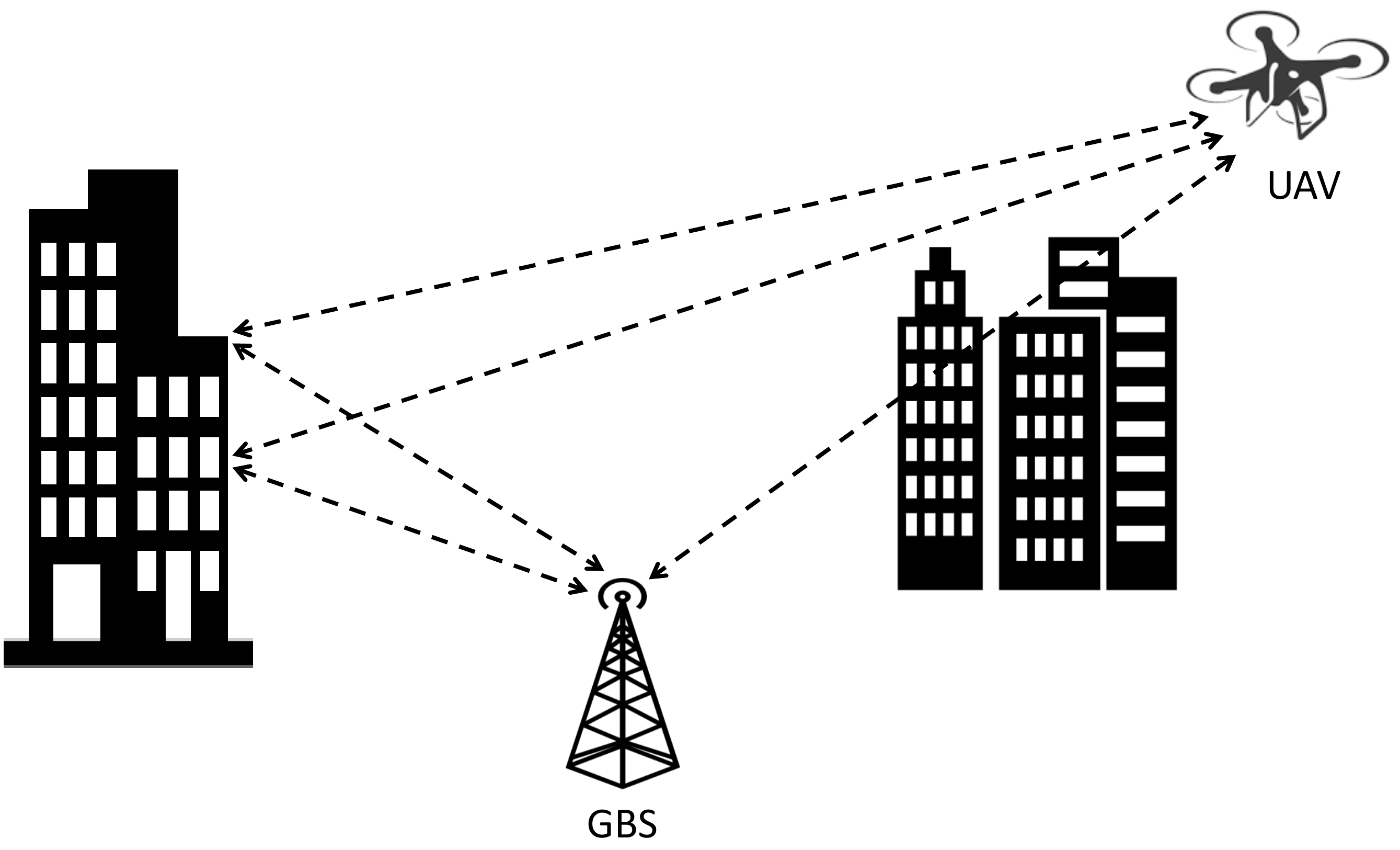}
	\vspace{-3mm}
	\caption{Cellular-connected UAV communication subjected to large-scale path loss and shadowing.}\label{RM_channel}
	\vspace{-6mm}
\end{figure}

Compared to traditional cellular communications serving the terrestrial users, new challenges arise in cellular-connected UAV communications. Specifically, with \emph{high flying altitude}, UAVs usually possess strong channels dominated by the \emph{line-of-sight (LoS)} paths with a much larger number of GBSs compared to terrestrial users, which leads to enhanced macro-diversity but also causes more severe co-channel interference with terrestrial communications. This has motivated several recent works on exploiting the UAV macro-diversity for cooperative processing by GBSs to deal with the strong aerial-ground interference problem (see, e.g., \cite{multibeam,Mei2,Howto}). Moreover, another unique characteristic of the UAV is its \emph{flexible mobility} over the three-dimensional (3D) space. This renders the UAV's \emph{trajectory} or \emph{path} an important new design parameter for improving its communication performance by proactively creating favorable channels with its associated GBSs via offline/online trajectory or path optimization/adaptation, which has been recently investigated in e.g., \cite{cellularUAV_arXiv,trajectoryoutage,Disconnectivity,Reinforcement}. In particular, trajectory design or path planning for cellular-connected UAV is usually performed \emph{offline} prior to the UAV's flight based on the mission requirement (e.g., flight time, initial/final locations) and available \emph{channel knowledge} with the GBSs at known locations in the UAV's fly region. For example, for rural areas without large obstacles above the GBSs, the GBS-UAV channels can be modeled as \emph{LoS}, based on which the UAV trajectory/path optimization problems subject to various communication constraints have been studied in \cite{cellularUAV_arXiv,trajectoryoutage,Disconnectivity}. However, the LoS air-ground channel model is not accurate for urban/suburban environments when the UAV's altitude is not sufficiently high, where the \emph{shadowing and multi-path fading effects} become non-negligible due to signal blockage and reflection/diffraction by e.g., large-size obstacles such as buildings, as illustrated in Fig. \ref{RM_channel}. In this case, more sophisticated channel models such as elevation angle-dependent \emph{Rician fading} \cite{URLLCPollin} and \emph{probabilistic LoS} models \cite{LAP} have been proposed, and offline UAV trajectory optimization based on such statistical channel models has been studied in \cite{You,Map_based}. It is worth noting that such statistical channel based UAV trajectory designs can only ensure the UAV communication performance on an average sense, while the actual performance at each location along its trajectory cannot be guaranteed in general due to the lack of location-specific channel knowledge.

Motivated by the above, we present in this paper a new \emph{radio map} based approach for UAV path planning in the cellular network. In general, radio maps contain rich information on the spectral activities and/or propagation channels over space and frequency, by averaging over the small-scale channel fading and its induced effects (e.g., power control) \cite{engineering}. In this paper, we refer the radio map specifically to the \emph{``large-scale channel gain map''} that contains information on the large-scale channel gains between each GBS and uniformly sampled locations on a 3D grid over the region of interest; and for convenience we use the above two terms interchangeably in the sequel of this paper. Note that in practice, such radio maps for GBSs can be obtained offline by deploying dedicated UAVs for channel sounding and measurements \cite{engineering}. Based on the given radio maps, we investigate the offline path planning for a cellular-connected UAV to minimize its flying distance (or time duration with a given constant speed) between a given pair of initial and final locations, subject to that the UAV needs to ensure a target link quality in terms of the large-scale channel gain with each of its associated GBSs at every time instant during the flight. We first transform this problem into an equivalent \emph{shortest path problem (SPP)} in graph theory, based on which the optimal solution is obtained via the Dijkstra algorithm \cite{graph}. Moreover, to reduce the computation complexity for finding the optimal SPP solution, we devise a grid quantization method that conducts a more coarse sampling of the radio map, by exploiting the potential spatial channel correlation among neighboring grid points. Based on this method, we solve approximate SPPs over reduced-size graphs to obtain suboptimal solutions with lower complexity. Numerical results validate the effectiveness of the proposed solutions in minimizing the path distance under the communication quality constraint. In addition, it is shown that the radio map quantization ratio can be adjusted to achieve a flexible trade-off between performance and complexity.
\vspace{-1mm}
\section{System Model}\label{sec_system}
Consider a cellular-connected UAV and $M\geq 1$ GBSs that may potentially be associated with the UAV during its flight. The UAV has a mission of flying from an initial location $U_0$ to a final location $U_F$, while communicating with one of the $M$ GBSs during the flight. We consider a 3D Cartesian coordinate system, where we denote $\tilde{\mv{u}}_0=[x_0,y_0,H_0]^T$ and $\tilde{\mv{u}}_F=[x_F,y_F,H_F]^T$ as the coordinates of $U_0$ and $U_F$, respectively; $\tilde{\mv{g}}_m=[a_m,b_m,H_{\mathrm{G}}]^T$ as the coordinate of each $m$th GBS, with $H_{\mathrm{G}}$ denoting the GBS height, which is assumed to be equal for all GBSs; and $\tilde{\mv{u}}(t)=[x(t),y(t),H(t)]^T,\ 0\leq t\leq T$ as the time-varying coordinate of the UAV, with $T$ denoting the mission completion time. We assume that the UAV flies at a constant speed denoted as $V$ meter/second (m/s), thus the UAV's trajectory $\{\tilde{\mv{u}}(t),0\leq t\leq T\}$ is determined solely by its flying path. Moreover, for ease of exposition, we assume that both the UAV and each GBS are equipped with an isotropic antenna with unit gain, while our results can be readily extended to the case with multiple antennas and/or other antenna patterns.

As illustrated in Fig. \ref{RM_channel}, the {large-scale channel gain} between each GBS and the UAV is the combination of the distance-dependent path loss and the shadowing, which are generally dependent on the locations of the GBS and UAV. Moreover, {small-scale fading} is also present in the GBS-UAV channels due to random/moving scatters on the ground. Without loss of generality, let $h_m(\tilde{\mv{u}})=\bar{h}_m(\tilde{\mv{u}})\tilde{h}_m(\tilde{\mv{u}})$ denote the instantaneous channel gain between each $m$th GBS and the UAV at location $\tilde{\mv{u}}$, where $\bar{h}_m(\tilde{\mv{u}})$ denotes the large-scale channel gain and $\tilde{h}_m(\tilde{\mv{u}})$ denotes the small-scale fading gain with normalized average power, i.e., $\mathbb{E}[\tilde{h}_m^2(\tilde{\mv{u}})]=1$.

We assume that the UAV is associated with GBS indexed by $I(t)\in \mathcal{M}$ at time instant $t$ during its mission, where $\mathcal{M}=\{1,...,M\}$. For both downlink and uplink communications, the receive signal-to-interference-plus-noise ratio (SINR) at each time instant $t$ can be modeled as
\vspace{-1mm}\begin{equation}\label{SINR}
\gamma(t)=\frac{P{h}^2_{I(t)}(\tilde{\mv{u}}(t))}{\sigma^2(t)}=\frac{P\bar{h}_{I(t)}^2(\tilde{\mv{u}}(t))\tilde{h}^2_{I(t)}(\tilde{\mv{u}}(t))}{\sigma^2(t)},
\vspace{-1mm}\end{equation}
where $P$ denotes the transmission power; $\sigma^2(t)$ denotes the interference-plus-noise power at the receiver. Note that the received interference power is subjected to the small-scale channel fading from other co-channel GBSs/terrestrial users and thus varies in a similar time scale as $\tilde{h}_{I(t)}(\tilde{\mv{u}}(t))$.
\begin{figure*}[t]
	\centering
	\subfigure[GBS and obstacle locations]{
	\includegraphics[width=4.4cm]{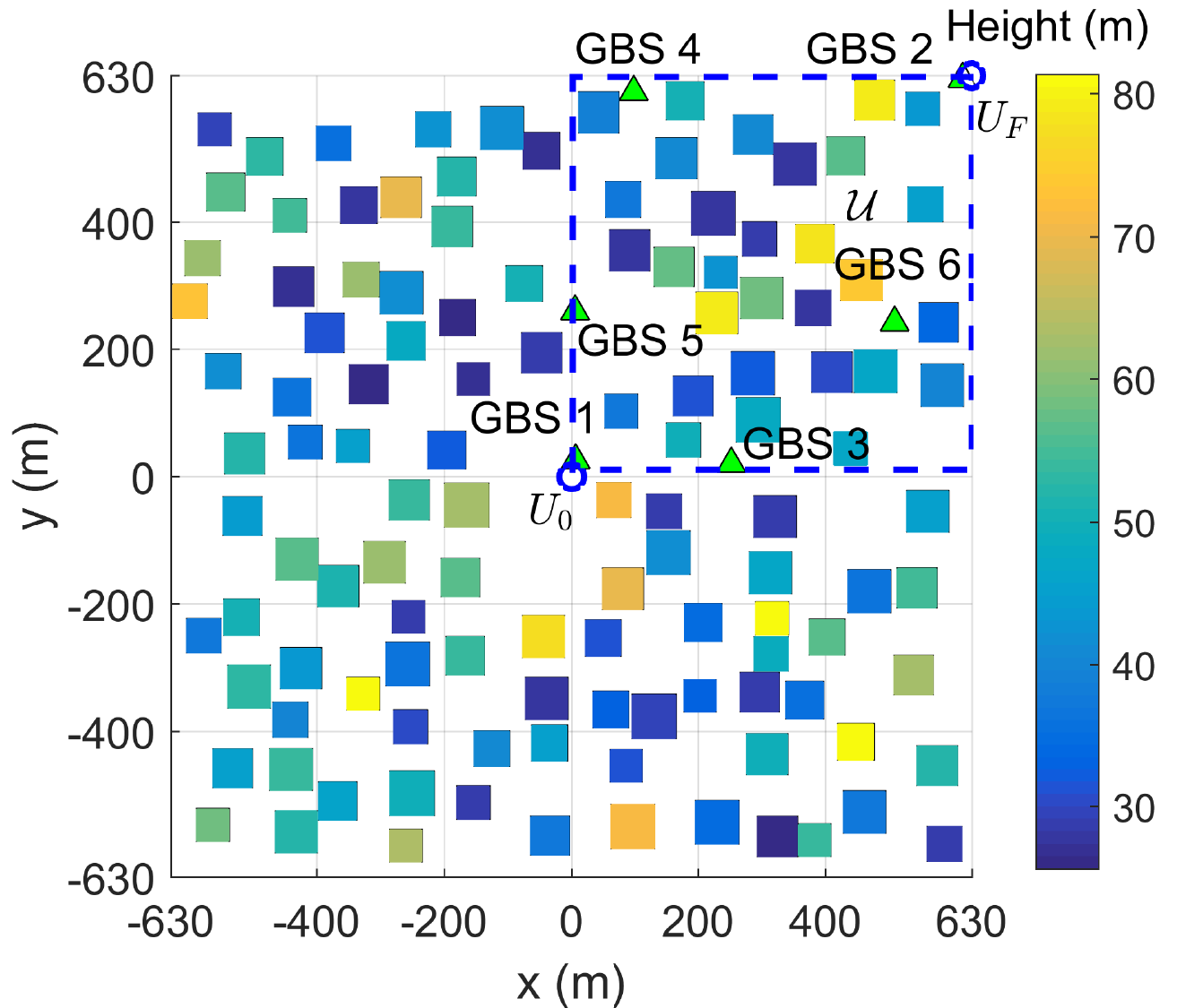}}
	\hspace{-3mm}
	\subfigure[Radio map for GBS 1]{
	\includegraphics[width=4.45cm]{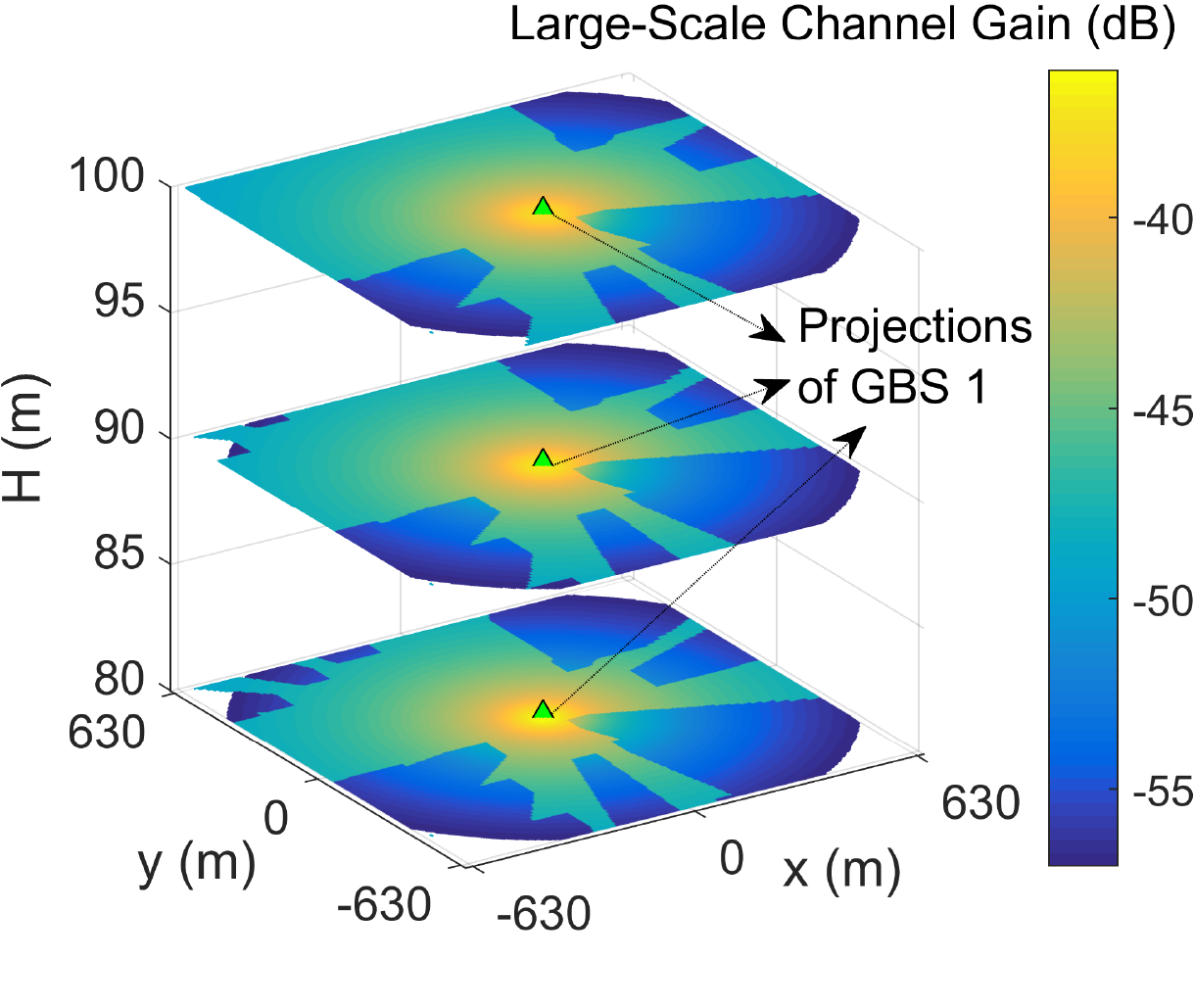}}
	\hspace{-3mm}
	\subfigure[Radio map for GBS 1 over $\mathcal{U}$]{
	\includegraphics[width=4.4cm]{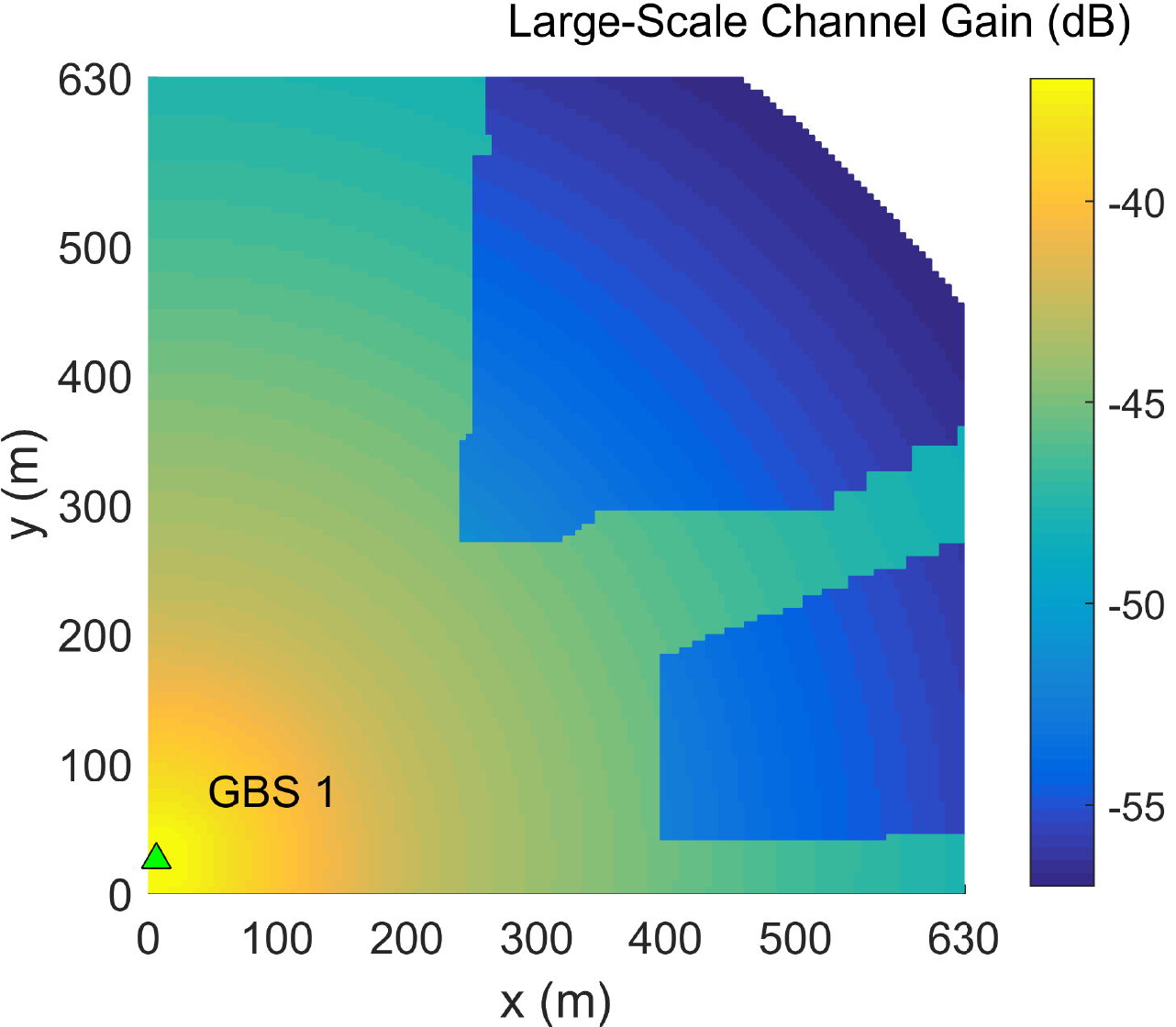}}
	\hspace{-3mm}
	\subfigure[Superposed radio map over $\mathcal{U}$]{
	\includegraphics[width=4.4cm]{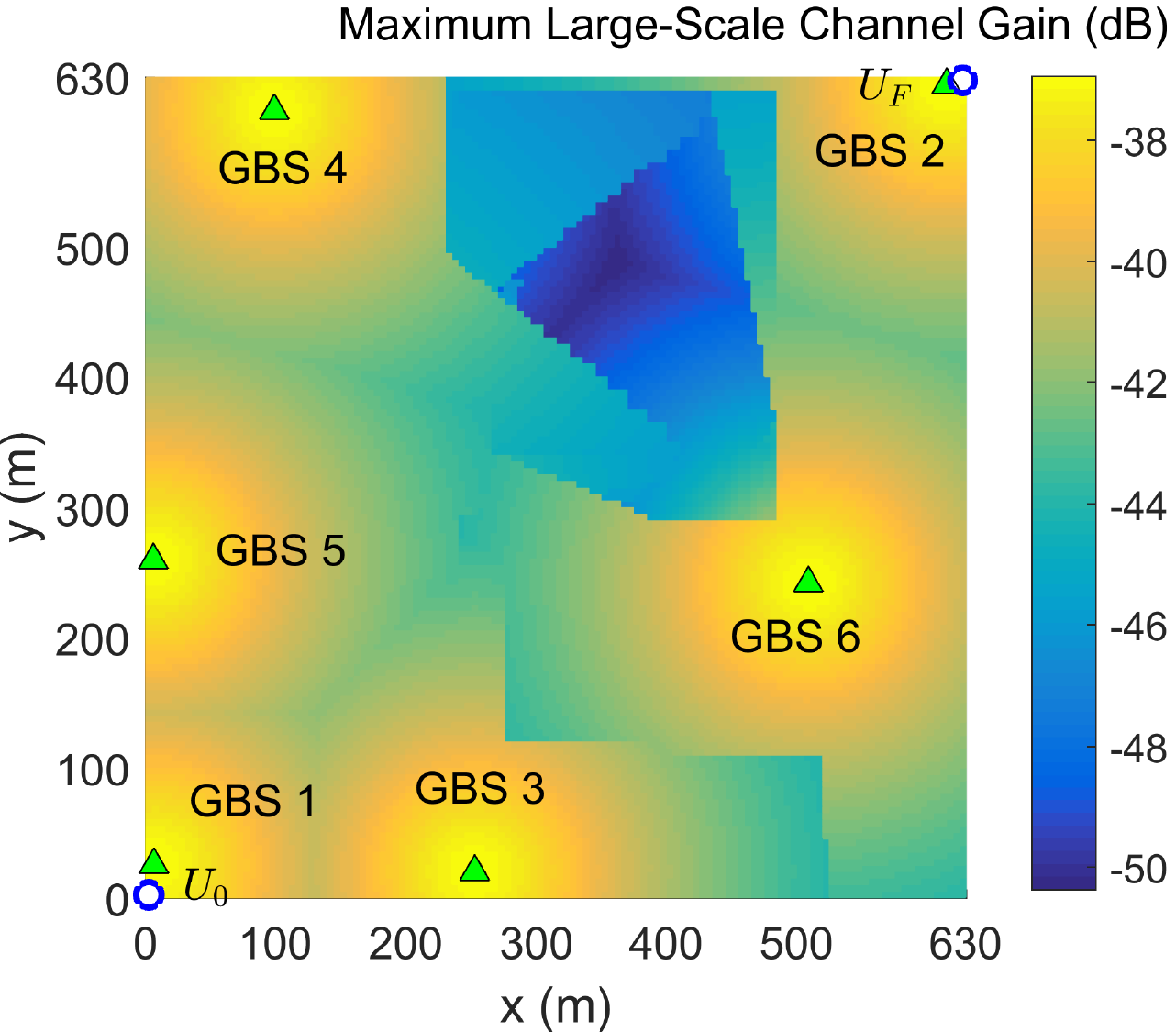}}
	\vspace{-2mm}
	\caption{Illustration of radio map.}\label{Radio_Map}
	\vspace{-6mm}
\end{figure*}

In contrast to $\tilde{h}_{I(t)}(\tilde{\mv{u}}(t))$ and $\sigma^2(t)$ which change fast over channel coherence time, the large-scale channel gain $\bar{h}_{I(t)}(\tilde{\mv{u}}(t))$ is determined by the locations, shapes, and dielectric properties of the large and high obstacles (e.g., buildings) that are generally static, and is thus practically constant for given UAV location $\tilde{\mv{u}}(t)$ and associated GBS $I(t)$. As such, such large-scale channel gains over different UAV locations with each GBS can be measured offline and stored in the so-called \emph{radio map}, for which the details will be given in Section \ref{sec_radio}. Motivated by the above, we adopt the \emph{large-scale channel gain} $\bar{h}_{I(t)}(\tilde{\mv{u}}(t))$ as the communication performance metric for UAV offline path planning, which specifies the expected quality of GBS-UAV communications when the UAV is located at $\tilde{\mv{u}}(t)$; while the impairments of small-scale fading and time-varying interference can be dealt with online via countermeasures such as channel coding and power control. We further assume a target on the large-scale channel gain denoted by $\bar{h}$, which needs to be achieved throughout the UAV's flight to meet the minimum link quality required for its mission (e.g., flight information exchange, video uploading, and so on), namely,
\vspace{-1mm}\begin{equation}\label{target}
\bar{h}_{I(t)}(\tilde{\mv{u}}(t))\geq \bar{h},\quad 0\leq t\leq T.
\vspace{-1mm}\end{equation}
To satisfy (\ref{target}), it is optimal to select the GBS that has the maximum large-scale channel gain to serve the UAV at time $t$, i.e., $I(t)=\arg\underset{m\in \mathcal{M}}{\max}\ \bar{h}_m(\tilde{\mv{u}}(t))$. Consequently, (\ref{target}) can be equivalently rewritten as follows:
\vspace{-1mm}\begin{equation}\label{channelconstraint}
\underset{m\in \mathcal{M}}{\max}\ \bar{h}_m(\tilde{\mv{u}}(t))\geq \bar{h},\quad 0\leq t\leq T.
\vspace{-1mm}\end{equation}

In the next, we first introduce the characterization of large-scale channel gains using radio maps; then, we formulate and solve the UAV path planning problem under the constraint in (\ref{channelconstraint}) based on the radio map.
\vspace{-1mm}
\section{Radio Map}\label{sec_radio}
The radio map for each $m$th GBS refers to the spatial distribution of its large-scale channel gain over the 3D space, i.e., $\bar{h}_m(\tilde{\mv{u}})$'s with UAV at locations $\tilde{\mv{u}}\in \mathbb{R}^{3\times 1}$. To facilitate efficient storage in practice, the radio map for each $m$th GBS is typically depicted only for neighborhood locations with \emph{non-negligible} large-scale channel gains above a given minimum threshold denoted by $\epsilon$, so as to reduce the map size. Moreover, it is generally \emph{discretized} with a finite granularity $\Delta_D$, and thus each GBS's radio map can be efficiently represented by a 3D \emph{matrix} of finite size denoted by $\tilde{\bar{\mv{H}}}_m\in \mathbb{R}_+^{\tilde{X}\times \tilde{Y}\times \tilde{H}}$, in which each $(\tilde{i},\tilde{j},\tilde{k})$-th element represents the large-scale channel gain between the $m$th GBS and the $(\tilde{i},\tilde{j},\tilde{k})$-th location in the discretized 3D space 
with granularity $\Delta_D$. The size of $\tilde{\bar{\mv{H}}}_m$ specified by $\tilde{X}$, $\tilde{Y}$ and $\tilde{H}$ is determined by the number of discretized 3D locations that yield large-scale channel gains no smaller than $\epsilon$, while for simplicity, we assume $\bar{h}_m(\tilde{\mv{u}})=0$ for $\tilde{\mv{u}}$ that is outside the locations considered in $\tilde{\bar{\mv{H}}}_m$. For example, for an area with GBS and obstacle locations given in Fig. \ref{Radio_Map} (a), where $H_{\mathrm{G}}=25$ m, Fig. \ref{Radio_Map} (b) shows the radio map for GBS 1 at altitude $H=80$ m, $90$ m, and $100$ m, respectively, with $\epsilon=-57$ dB and $\Delta_D=5$ m, for which the channel parameters will be given later in Section \ref{sec_num}. It can be observed that in such a dense urban environment, the large-scale channel gain behaves drastically different from that under the LoS channel model, where two different locations with equal distance to the same GBS may have significantly different gains due to heterogeneous shadowing effects.

Furthermore, for UAV path planning with given initial/final locations, we only need to consider radio maps of GBSs that overlap with a target region that is sufficiently large to cover all possible UAV locations during its flight. For example, we assume in this paper that the UAV flies at a \emph{fixed altitude} with $H(t)=H,0\leq t\leq T$, where $H$ is no smaller than the maximum obstacle height to avoid collision. We will extend our results to the more general 3D flight case in the journal version of this work. As such, we only need to consider the UAV's horizontal locations during its flight, i.e., $\{{\mv{u}}(t)=[x(t),y(t)]^T,0\leq t\leq T\}$, in a two-dimensional (2D) square region denoted by $\mathcal{U}\subset \mathbb{R}^{2\times 1}$ with edge length $L$, which is chosen to be sufficiently large to cover all possible UAV horizontal locations during the flight. Note that $L$ increases with the distance between initial and final UAV locations $U_0$ and $U_F$. The discretized UAV locations in $\mathcal{U}$ form a $D\times D$ grid with granularity $\Delta_D$ and $D=L/\Delta_D$, denoted as $\mathcal{U}_D=\{{\mv{u}}_D(i,j):i\in {\mathcal{D}},j\in {\mathcal{D}}\}$, with $\mathcal{D}=\{1,...,D\}$ and ${\mv{u}}_D(i,j)$ \hbox{denoting the $(i,j)$-th location on the grid, which is given by}
\vspace{-1mm}\begin{equation}
{\mv{u}}_D(i,j)=[i-1/2,j-1/2]^T\Delta_{{D}},\quad i,j\in {\mathcal{D}}.
\vspace{-1mm}\end{equation} 
The (effective) radio map for each $m$th GBS over $\mathcal{U}$ can be thereby represented by a 2D {matrix} $\bar{\mv{H}}_m\in \mathbb{R}_+^{D\times D}$, where each $(i,j)$-th element $[\bar{\mv{H}}_m]_{i,j}$ denotes the large-scale channel gain between GBS $m$ and $\mv{u}_D(i,j)$, namely,
\vspace{-1mm}\begin{equation}\label{map_matrix}
[\bar{\mv{H}}_m]_{i,j}=\bar{h}_m([\mv{u}_D(i,j)^T,H]^T),\quad m\in \mathcal{M},\ i,j\in \mathcal{D}.
\vspace{-1mm}\end{equation}
Note that the size of the above radio map matrices over $\mathcal{U}$ is different from that of the original radio map matrices $\tilde{\bar{\mv{H}}}_m$'s due to map truncation and expansion with zero-padding (for locations in $\mathcal{U}$ that do not overlap with each $\tilde{\bar{\mv{H}}}_m$). For illustration, with an example of $\mathcal{U}$ given in Fig. \ref{Radio_Map} (a), we show in Fig. \ref{Radio_Map} (c) the radio map for GBS 1 over $\mathcal{U}$ with $H=90$ m, which is obtained according to $\tilde{\bar{\mv{H}}}_1$. In addition, based on (\ref{channelconstraint}), we depict in Fig. \ref{Radio_Map} (d) the so-called \emph{superposed radio map} over $\mathcal{U}$ showing the maximum large-scale channel gain at each $(i,j)$-th grid point among all the GBSs whose radio maps overlap with $\mathcal{U}$, i.e., $\underset{m\in \mathcal{M}}{\max}\ [\bar{\mv{H}}_m]_{i,j}$, which can be observed to vary more abruptly than the radio maps for individual GBSs. In the sequel, we assume that the radio maps for the $M$ GBSs, $\{\bar{\mv{H}}_m\}_{m=1}^M$, are \emph{perfectly} known with granularity $\Delta_D$, which is \emph{sufficiently small} such that $\bar{h}_m([{\mv{u}}^T,H]^T)=[\bar{\mv{H}}_m]_{i,j}$ holds for any UAV horizontal location $\mv{u}$ in the $(i,j)$-th grid \emph{cell} (i.e., $\mv{u}$ that satisfies $|{\mv{u}}-\mv{u}_D(i,j)|\preceq\frac{\Delta_D}{2}[1,1]^T$, with $\mv{a}\preceq \mv{b}$ denoting that $\mv{a}$ is element-wise no larger than $\mv{b}$, as illustrated in Fig. \ref{grid1} (a)). In other words, the radio maps $\{\bar{\mv{H}}_m\}_{m=1}^M$ are able to characterize the large-scale channel gain distributions with sufficiently high accuracy.
\vspace{-1mm}
\section{Problem Formulation}
Based on the given radio maps of the $M$ GBSs, $\{\bar{\mv{H}}_m\}_{m=1}^M$ in (\ref{map_matrix}), we aim to minimize the UAV's flying distance from $U_0$ to $U_F$ by optimizing its (horizontal) path over the 2D grid $\mathcal{U}_D$ subject to the large-scale channel gain constraint given in (\ref{channelconstraint}) for all the grid points along the UAV path. To this end, we need to find a UAV path consisting of a sequence of \emph{connected line segments}, where two end points of each segment are \emph{adjacent grid points} from $\mathcal{U}_D$ with distance $\Delta_D$ or $\sqrt{2}\Delta_D$, as illustrated in Fig. \ref{grid1} (a). This is motivated by the fact that if two adjacent grid points both satisfy the large-scale channel gain constraint, any point $\mv{u}$ on the line segment between them also satisfies this constraint. For convenience, we assume that the initial and final horizontal locations of the UAV given by $\mv{u}_0=[x_0,y_0]^T$ and $\mv{u}_F=[x_F,y_F]^T$ are both on the grid $\mathcal{U}_D$. Therefore, we formulate the following optimization problem:
\vspace{-1mm}\begin{align}
\mbox{(P1)}\underset{K,\{i_k,j_k\}_{k=1}^K}{\min} &\sum_{k=1}^{K-1} \|{\mv{u}}_{D}(i_{k+1},j_{k+1})-{\mv{u}}_D(i_k,j_k)\|\\[-1mm]
\mathrm{s.t.}\qquad &\underset{m\in \mathcal{M}}{\max}\ [\bar{{\mv{H}}}_m]_{i_k,j_k}\geq \bar{h},\quad k=1,...,K\label{P1c1}\\[-1mm]
& {\mv{u}}_D(i_1,j_1)={\mv{u}}_0\label{P1c2}\\[-1mm]
& \|{\mv{u}}_D(i_{k+1},j_{k+1})-{\mv{u}}_D(i_k,j_k)\|\leq \sqrt{2}{\Delta}_D,\nonumber\\[-1mm]
&\qquad\qquad\qquad\qquad k=1,...,K-1\\[-1mm]
& {\mv{u}}_D(i_K,j_K)={\mv{u}}_F\label{P1c4}\\[-1mm]
& i_k,j_k\in \mathcal{D},\quad k=1,...,K,
\end{align}
where $\|\cdot\|$ denotes the Euclidean norm, and $K$ denotes the number of grid points that the UAV traverses over its flight. Note that (P1) is a non-convex combinatorial optimization problem due to the integer variables $\{i_k,j_k\}_{k=1}^K$ and $K$. Thus, it cannot be solved efficiently via standard optimization methods. In the following, we reformulate (P1) based on graph theory, and propose both the optimal and low-complexity suboptimal solutions for it.
\vspace{-1mm}
\section{Optimal Solution}\label{sec_opt}
\vspace{-1mm}
In this section, we obtain the optimal solution to (P1) by casting it as an equivalent \emph{SPP} in graph theory \cite{graph}. To this end, a straightforward approach is to consider all the $D^2$ grid points in $\mathcal{U}_D$ in the vertex set of an equivalent graph. However, this may be inefficient since under the constraints in (\ref{P1c1}), only the $({i},{j})$-th grid points with $\underset{m\in \mathcal{M}}{\max}\ [\bar{{\mv{H}}}_m]_{{i},{j}}\geq\bar{h}$ may potentially constitute a feasible path. Thus, we first consider the following \emph{radio map preprocessing} to identify these grid points, which are referred to as the \emph{``feasible grid points''}.
\vspace{-1mm}
\subsection{Radio Map Preprocessing}
\vspace{-1mm}
Specifically, we construct a new \emph{``feasible map''} denoted by ${\mv{F}}\in \{0,1\}^{D\times D}$ based on $\{\bar{\mv{H}}_m\}_{m=1}^M$, where each $(i,j)$-th element is given by
\vspace{-1mm}\begin{equation}\label{feasible}
[\mv{F}]_{i,j}=\begin{cases}
1,\ \mathrm{if}\ \underset{m\in \mathcal{M}}{\max}\ [\bar{\mv{H}}_m]_{i,j}\geq \bar{h}\\[-1mm]
0,\ \mathrm{otherwise},
\end{cases} i,j\in \mathcal{D}.
\vspace{-1mm}\end{equation}
Note that $[\mv{F}]_{i,j}=1$ indicates that the $(i,j)$-th grid point is a feasible grid point, and $[\mv{F}]_{i,j}=0$ otherwise. The complexity for the above preprocessing can be shown to be $\mathcal{O}(D^2M)$.
\vspace{-1mm}\subsection{Graph Based Problem Reformulation and Solution}\vspace{-1mm}
Next, based on the constructed feasible map ${\mv{F}}$, we propose an equivalent graph based reformulation of (P1). Specifically, we construct an undirected weighted graph $G_{\mathrm{D}}=(V_{\mathrm{D}},E_{\mathrm{D}})$ \cite{graph}. The vertex set of $G_{\mathrm{D}}$ is given by
\vspace{-1mm}\begin{equation}
V_{\mathrm{D}}=\{U_D(i,j):[{\mv{F}}]_{i,j}=1,i\in \mathcal{D},j\in \mathcal{D}\},\!
\vspace{-1mm}\end{equation}
where $U_{{D}}(i,j)$ represents the $(i,j)$-th (feasible) grid point with location ${\mv{u}}_D(i,j)$. The edge set of $G_{\mathrm{D}}$ is given by
\vspace{-1mm}\begin{align}
E_{\mathrm{D}}\!\!=\{&(U_D(i,j),U_D(i',j'))\!:\nonumber\\
&\|{\mv{u}}_D(i,j)\!-\!{\mv{u}}_D(i',j')\|\!\leq\! \sqrt{2}{\Delta}_D\}.
\end{align}
Note that an edge exists between two vertices $U_D(i,j)$ and $U_D(i',j')$ if and only if the corresponding two grid points are adjacent. Furthermore, the weight of each edge is given by
\vspace{-1mm}\begin{equation}
W_{\mathrm{D}}(U_D(i,j),U_D(i',j'))=\|{\mv{u}}_D(i,j)-{\mv{u}}_D(i',j')\|,
\vspace{-1mm}\end{equation}
which represents the flying distance between the two corresponding locations.

With graph $G_{\mathrm{D}}$ constructed above, (P1) can be shown to be equivalent to finding the \emph{shortest path} from $U_D(i_1,j_1)$ to $U_D(i_K,j_K)$ in $G_{\mathrm{D}}$, where $(i_1,j_1)$ and $(i_K,j_K)$ are given in (\ref{P1c2}) and (\ref{P1c4}). This problem can be solved via the Dijkstra algorithm with worst-case complexity $\mathcal{O}(|E_{\mathrm{D}}|+|V_{\mathrm{D}}|\log|V_{\mathrm{D}}|)=\mathcal{O}(D^2\log D)$ using the Fibonacci heap structure \cite{graph}, where the worst-case values for $|E_{\mathrm{D}}|$ and $|V_{\mathrm{D}}|$ can be shown to be $2(D-1)(2D-1)$ and $D^2$, respectively. With the obtained shortest path denoted by $(U_D(i_1^\star,j_1^\star),...,U_D(i_{K^\star}^\star,j_{K^\star}^\star))$, the optimal solution to (P1) is obtained as $K^\star$ and $\{i_k^\star,j_k^\star\}_{k=1}^{K^\star}$. Note that the feasibility of (P1) is automatically checked via the Dijkstra algorithm, where (P1) is infeasible if and only if no path is returned by the algorithm. For illustration, we show in Fig. \ref{grid} the optimal solution for a 2D grid with the given feasible map. It can be shown that constructing the graph $G_{\mathrm{D}}$ requires worst-case complexity of $\mathcal{O}(D^2)$. Therefore, the overall worst-case complexity for finding the optimal solution to (P1) based on the feasible map $\mv{F}$ is $\mathcal{O}(D^2\log D)$.
\begin{figure}[t]
	\centering
	\includegraphics[width=8cm]{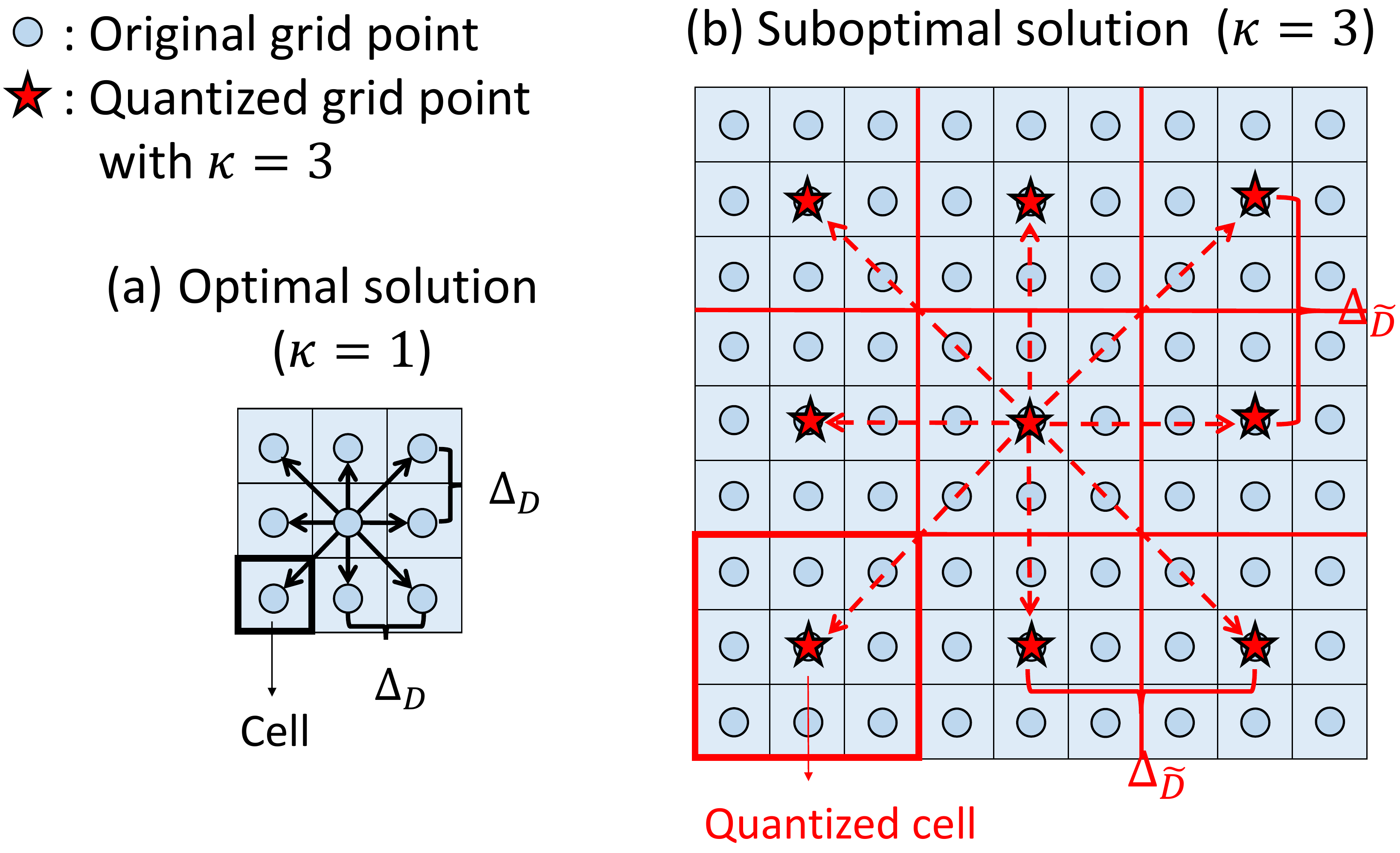}
	\vspace{-3mm}
	\caption{Illustration of grid and path structures in proposed solutions to (P1).}\label{grid1}\vspace{-6mm}
\end{figure}
\vspace{-1mm}\section{Suboptimal Solution via Grid Quantization}\label{sec_sub}\vspace{-1mm}
Note that the \emph{complexity} for finding the optimal solution to (P1) scales up with $D$, while the value of $D=\frac{L}{\Delta_D}$ can be practically arbitrarily large with given $\Delta_D$ and increasing the edge length $L$ of the region of interest, $\mathcal{U}$. Moreover, it is worth noting that the required \emph{memory} for storing all the edge weights in graph $G_{\mathrm{D}}$ is dependent on $|E_{\mathrm{D}}|$, which also increases with $D$. For example, with $\Delta_D=5$ m and $L=100$ km when $U_0$ and $U_F$ are far apart, we have $D=2\times 10^4$ and consequently $D^2\log D\approx 4\times 10^9$; in addition, we have $|E_{\mathrm{D}}|\approx 1.6\times 10^9$ in the worst-case, which demands for approximately $12.8$ GB memory for storing $G_{\mathrm{D}}$. In practice, such high complexity and large memory size are prohibitive or even unaffordable. To tackle this issue, we propose to reduce the number of vertices involved in the SPP by applying a \emph{grid quantization method} and considering a new path structure composed of connected line segments between \emph{quantized grid points}, as specified in the rest of this section.
\vspace{-1mm}\subsection{Radio Map Preprocessing}\vspace{-1mm}
To start with, we present the proposed grid quantization method. Denote $\kappa\in \mathbb{N}_+$ as the \emph{quantization ratio}, with $\kappa\geq 1$ and $\mathbb{N}_+$ denoting the set of positive integers. By applying \emph{uniform quantization} over the grid points in $\mathcal{U}_D$, we obtain $\tilde{D}^2$ points with granularity $\Delta_{\tilde{D}}$, where $\tilde{D}=D/\kappa\leq D$ and $\Delta_{\tilde{D}}=\kappa\Delta_D\geq \Delta_D$. For ease of exposition, we assume $D/\kappa$ is an integer, and $\kappa$ is an odd number. Let $\mathcal{U}_{\tilde{D}}=\{{\mv{u}}_{\tilde{D}}(i,j):i\in \tilde{\mathcal{D}},j\in \tilde{\mathcal{D}}\}$ denote the \emph{quantized grid}, with $\tilde{\mathcal{D}}=\{1,...,\tilde{D}\}$ and ${\mv{u}}_{\tilde{D}}(i,j)$ denoting the $(i,j)$-th location on the quantized grid, which is given by
\vspace{-1mm}\begin{equation}
{\mv{u}}_{\tilde{D}}(i,j)=[i-1/2,j-1/2]^T\Delta_{\tilde{D}}, \quad i,j\in\tilde{\mathcal{D}}.
\vspace{-1mm}\end{equation}
The proposed grid quantization method is illustrated in Fig. \ref{grid1} (b) for $\kappa=3$. Notice that each ${\mv{u}}_{\tilde{D}}(i,j)$ lies among $\kappa^2$ original grid points indexed by a \emph{``neighboring set''} $\mathcal{N}(i,j)$, whose corresponding cells form a \emph{``quantized cell''}, as illustrated in Fig. \ref{grid1} (b). Specifically, we have $\mathcal{N}(i,j)=\{(p,q):|{\mv{u}}_D(p,q)-{\mv{u}}_{\tilde{D}}(i,j)|\preceq \frac{\Delta_{\tilde{D}}}{2}[1,1]^T,p\in \mathcal{D},q\in \mathcal{D}\}$. This is motivated by the fact that the channels for neighboring grid points in $\mathcal{N}(i,j)$ are typically highly correlated, thus they can be ``well-represented'' by one single quantized grid point ${\mv{u}}_{\tilde{D}}(i,j)$ at the center.

\begin{figure}[t]
	\centering
	\includegraphics[width=8cm]{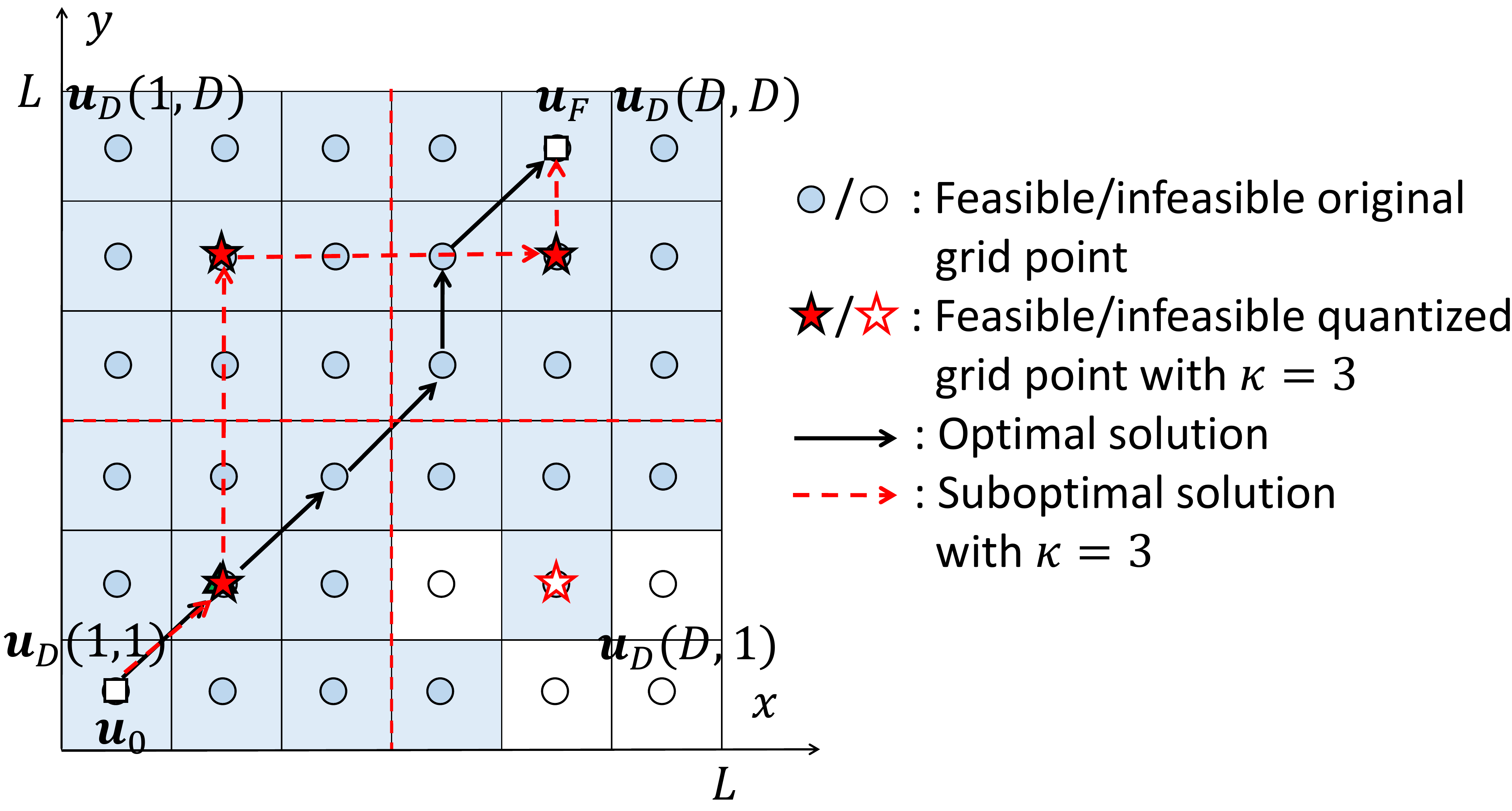}
	\vspace{-3mm}
	\caption{Illustration of the proposed path solutions to (P1).}\label{grid}
	\vspace{-6mm}
\end{figure}

Next, we consider a feasible path structure where ${\mv{u}}_{\tilde{D}}(i,j)$ is a \emph{``feasible quantized grid point''} (namely, may potentially constitute a feasible path) if and only if all the neighboring original grid points in $\mathcal{N}(i,j)$ are feasible grid points defined by (\ref{feasible}), such that its connected line segment with another adjacent feasible quantized grid point at any direction does not violate the large-scale channel gain constraint. To identify such points, we construct a new \emph{``quantized feasible map''} denoted by $\tilde{\mv{F}}\in \{0,1\}^{\tilde{D}\times \tilde{D}}$, where each $(i,j)$-th element is given by
\vspace{-1mm}\begin{equation}
\!\!\!\!\!\![\tilde{\mv{F}}]_{i,j}\!=\!\begin{cases}\!
1,\mathrm{if} \underset{m\in \mathcal{M}}{\max}[\bar{\mv{H}}_m]_{p,q}\!\geq\! \bar{h},\forall (p,q)\!\in\! \mathcal{N}(i,j)\\[-1mm]
\!0,\mathrm{otherwise},
\end{cases}\!\!\!\!\! i,j\!\in \!\tilde{\mathcal{D}}.
\vspace{-1mm}\end{equation}
Note that $[\tilde{\mv{F}}]_{i,j}=1$ indicates that the $(i,j)$-th quantized grid point is a feasible quantized grid point, and $[\tilde{\mv{F}}]_{i,j}=0$ otherwise. The complexity for the above preprocessing can be shown to be $\mathcal{O}(\tilde{D}^2\kappa^2M)=\mathcal{O}(D^2M)$.

\vspace{-1mm}\subsection{Reduced-Size Graph and Suboptimal Solution}\vspace{-1mm}
Based on the quantized feasible map $\tilde{\mv{F}}$, we can introduce the new path structure. Specifically, assuming that ${\mv{u}}_0$ and ${\mv{u}}_F$ belong to the $(\tilde{i}_1,\tilde{j}_1)$-th and $(\tilde{i}_K,\tilde{j}_K)$-th quantized cells, respectively, we first let the UAV fly from ${\mv{u}}_0$ to ${\mv{u}}_{\tilde{D}}(\tilde{i}_1,\tilde{j}_1)$ at the start, and from ${\mv{u}}_{\tilde{D}}(\tilde{i}_K,\tilde{j}_K)$ to ${\mv{u}}_F$ in the end. Moreover, we assume that a path exists between two feasible quantized grid points if and only if they are adjacent with distance $\Delta_{\tilde{D}}$ or $\sqrt{2}\Delta_{\tilde{D}}$. Thus, the UAV can fly from a feasible quantized grid point indexed by $(i,j)$ to adjacent points in $8$ directions, whose index set is given by $\mathcal{A}(i,j)=\{(k,h):\|{\mv{u}}_{\tilde{D}}(i,j)\!-\!{\mv{u}}_{\tilde{D}}(k,h)\|\!\leq\! \sqrt{2}{\Delta}_{\tilde{D}},k\in \tilde{\mathcal{D}},h\in \tilde{\mathcal{D}}\}$, as illustrated in Fig. \ref{grid1} (b). Under the above structure, we construct an undirected weighted graph $G_{\tilde{\mathrm{D}}}=(V_{\tilde{\mathrm{D}}},E_{\tilde{\mathrm{D}}})$ with vertex set
\vspace{-1mm}\begin{equation}
V_{\tilde{\mathrm{D}}}=\{U_{\tilde{D}}(i,j):[\tilde{\mv{F}}]_{i,j}\!=\!1,i\!\in\! \tilde{\mathcal{D}},j\!\in\! \tilde{\mathcal{D}}\},
\vspace{-1mm}\end{equation}
where $U_{\tilde{D}}(i,j)$ denotes the $(i,j)$-th (feasible) quantized grid point with location ${\mv{u}}_{\tilde{D}}(i,j)$. Note that $|V_{\tilde{\mathrm{D}}}|$ is significantly smaller than $|V_{{\mathrm{D}}}|$, with a worst-case value of $\tilde{D}^2=D^2/\kappa^2\leq D^2$. The edge set of $G_{\tilde{\mathrm{D}}}$ is given by
\vspace{-1mm}\begin{equation}\label{edgeI}
E_{\tilde{\mathrm{D}}}\!=\!\{(U_{\tilde{D}}(i,j),U_{\tilde{D}}(i',j')):(i',j')\in \mathcal{A}(i,j)\}.
\vspace{-1mm}\end{equation}
The weight of each edge is given by
\vspace{-1mm}\begin{equation}
W_{\tilde{\mathrm{D}}}(U_{\tilde{D}}(i,j),U_{\tilde{D}}(i',j'))=\|{\mv{u}}_{\tilde{D}}(i,j)-{\mv{u}}_{\tilde{D}}(i',j')\|.
\vspace{-1mm}\end{equation}
Note that (P1) under the proposed path structure is equivalent to finding the \emph{shortest path} from $U_{\tilde{D}}(\tilde{i}_1,\tilde{j}_1)$ to $U_{\tilde{D}}(\tilde{i}_K,\tilde{j}_K)$ in graph $G_{\tilde{\mathrm{D}}}$, which can be solved via the Dijkstra algorithm with worst-case complexity $\mathcal{O}(|E_{\tilde{\mathrm{D}}}|+|V_{\tilde{\mathrm{D}}}|\log|V_{\tilde{\mathrm{D}}}|)=\mathcal{O}(\tilde{D}^2\log \tilde{D})$ \cite{graph}. By noting that the construction of graph $G_{\tilde{\mathrm{D}}}$ requires worst-case complexity of $\mathcal{O}(\tilde{D}^2)$, the overall worst-case complexity for obtaining a suboptimal solution based on $\tilde{\mv{F}}$ is $\mathcal{O}(\tilde{D}^2\log \tilde{D})=\mathcal{O}((D/\kappa)^2\log (D/\kappa))$. An example of the proposed suboptimal solution is given in Fig. \ref{grid} with $\kappa=3$. It can be easily shown that the obtained shortest path in $G_{\tilde{\mathrm{D}}}$ always corresponds to a feasible solution to (P1) (as can be observed from Fig. \ref{grid}), which is optimal to (P1) with $\kappa=1$ (i.e., $\tilde{D}=D$), and generally suboptimal for $\kappa>1$. 

Finally, note that the overall complexities for the proposed optimal solution and suboptimal solution are given by $\mathcal{O}\left(D^2M+D^2\log D\right)$ and $\mathcal{O}(D^2M+(D/\kappa)^2\log(D/\kappa))$, respectively, which can be well-approximated by $\mathcal{O}(D^2\log D)$ and $\mathcal{O}((D/\kappa)^2\log D)$, respectively, for the practical case with $D\gg M$ and $D\gg \kappa$. Thus, the suboptimal solution only requires $1/\kappa^2$ of the complexity required by the optimal solution. Note that as $\kappa$ increases, the performance of the suboptimal solution generally degrades as the quantization becomes more coarse, while the required complexity also decreases. Thus, a flexible performance-complexity trade-off can be achieved by selecting the quantization ratio $\kappa$.

\vspace{-1mm}\section{Numerical Results}\label{sec_num}\vspace{-1mm}
In this section, we provide numerical results to evaluate the performance of our proposed path planning solutions. As illustrated in Fig. \ref{Radio_Map} (a), we consider a square area $\mathcal{U}$ with edge length $L=630$ m, over which $M=6$ GBSs are uniformly randomly distributed, each with height $H_{\mathrm{G}}=25$ m; moreover, $30$ obstacles are randomly distributed in $\mathcal{U}$, each modeled as a 3D cuboid with equal length and width randomly generated according to the uniform distribution in $[50,70]$ m, and height randomly generated according to the Rayleigh distribution with mean $40$ m, which is truncated to be no larger than the UAV's flying altitude set as $H=90$ m. The UAV's initial and final horizontal locations are set as ${\mv{u}}_0=[2.5,2.5]^T$ m and ${\mv{u}}_F=[627.5,627.5]^T$ m, respectively. The large-scale channel gain between each GBS and UAV location is modeled according to the 3GPP technical report based on the urban macro (UMa) scenario \cite{3GPPUAV}. The radio map granularity is set as $\Delta_D=5$ m, and the corresponding radio maps are illustrated in Fig. \ref{Radio_Map}.

\begin{figure}[t]
	\centering
	\includegraphics[width=7.5cm]{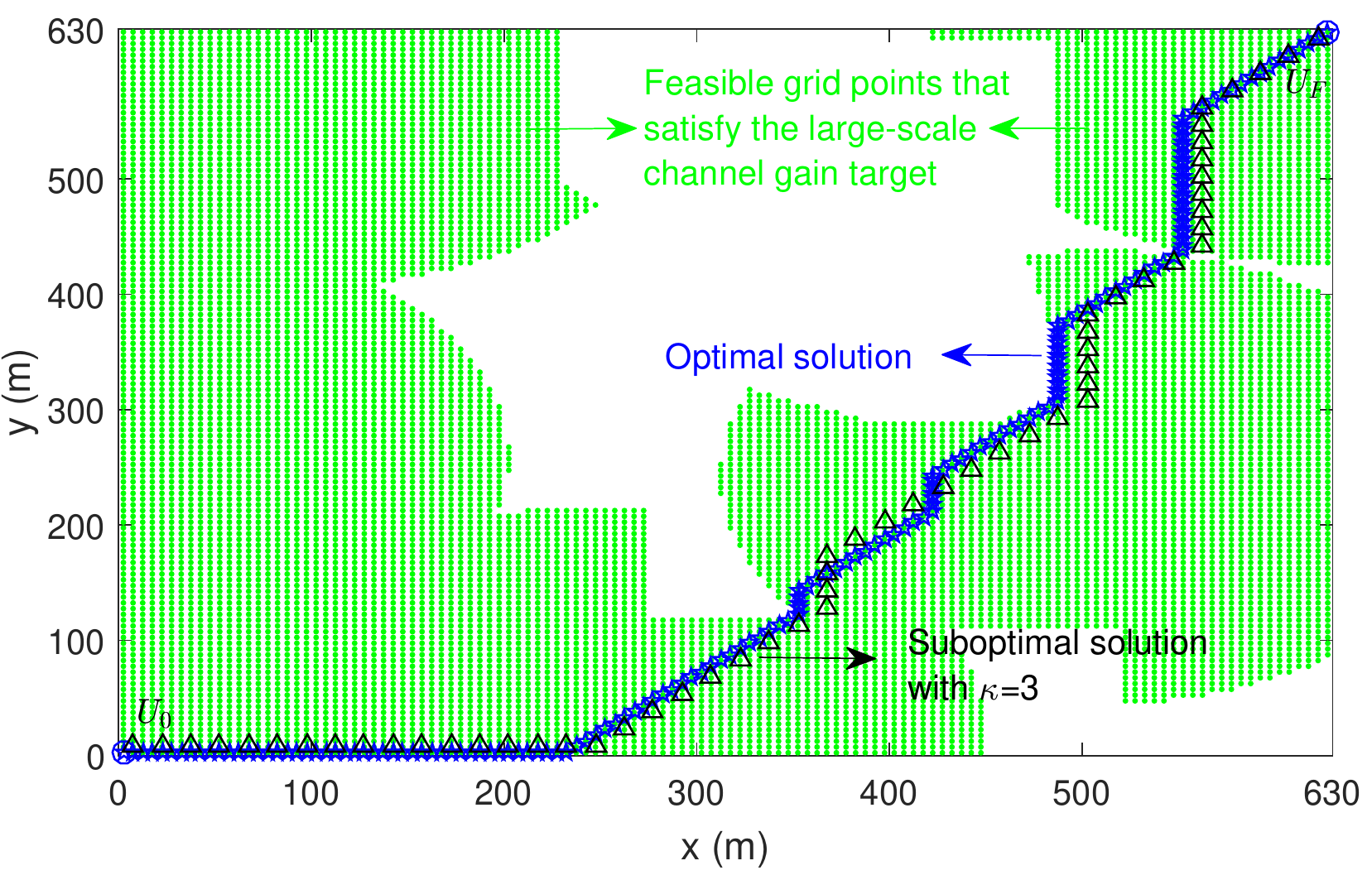}
	\vspace{-4mm}
	\caption{Illustration of the proposed path solutions with $\bar{h}=-42.5$ dB.}\label{path}
	\vspace{-4mm}	
\end{figure}

Under the above setup, we first consider a large-scale channel gain target $\bar{h}=-42.5$ dB and show in Fig. \ref{path} the proposed optimal solution and suboptimal solution with $\kappa=3$, where we also depict the feasible (original) grid points that satisfy the large-scale channel gain target for the purpose of illustration. It is observed that due to grid quantization, the suboptimal solution is less efficient (of longer path length) than the optimal solution.
\begin{figure}[t]
	\centering
	\includegraphics[width=7.5cm]{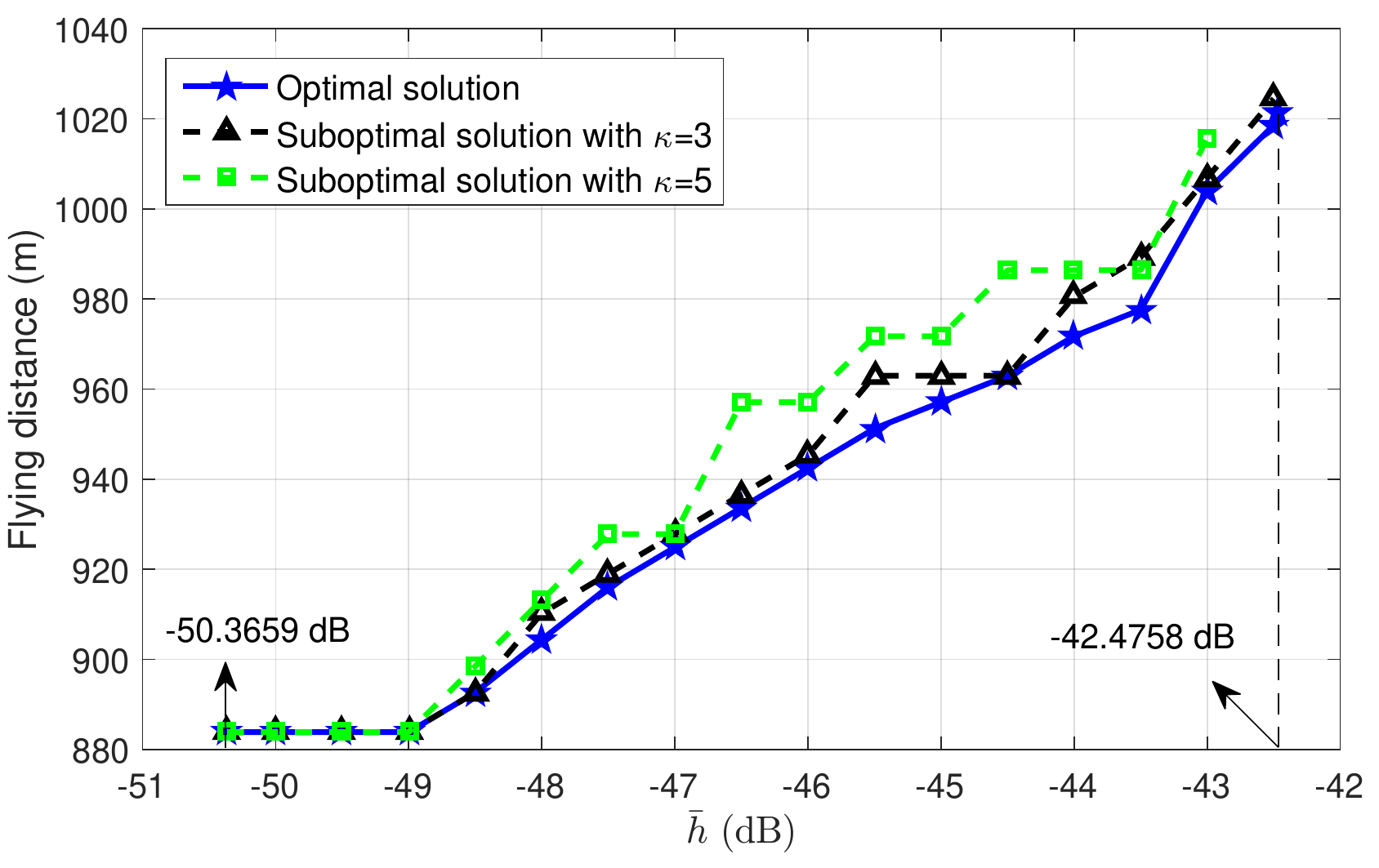}
	\vspace{-3mm}
	\caption{Flying distance versus $\bar{h}$ for proposed solutions.}\label{distance}
	\vspace{-7mm}	
\end{figure}

Next, we show in Fig. \ref{distance} the required flying distance from $U_0$ to $U_F$ with the proposed optimal and suboptimal solutions versus the large-scale channel gain target $\bar{h}$, where the quantization ratio for the suboptimal solution is set as $\kappa=3$ or $5$. Note that the minimum value of $\underset{m\in \mathcal{M}}{\max}\ [\bar{\mv{H}}_m]_{i,j}$ for all $i,j\in \mathcal{D}$ on the given map is $-50.3659$ dB; moreover, it is found via the Dijkstra algorithm that (P1) becomes infeasible if $\bar{h}>-42.4758$ dB for this setup. Thus, the range of $\bar{h}$ in Fig. \ref{distance} is set as $[-50.3659,-42.4758]$ dB. Note that under a practical setup with transmission power $P=23$ dBm, average interference-plus-noise power density $-150$ dBm/Hz with $7$ dB noise figure, and $10$ MHz bandwidth, this corresponds to an expected SINR target range of $[-4.7318,11.0484]$ dB. It is observed from Fig. \ref{distance} that the optimal solution is feasible for all values of $\bar{h}$, while the suboptimal solution with $\kappa=3$ and $\kappa=5$ becomes infeasible with $\bar{h}>-42.5$ dB and $\bar{h}>-43$ dB, respectively. Moreover, the required flying distance for the suboptimal solution generally increases as $\kappa$ increases, since the increasingly coarse grid quantization yields less flexibility in the path design, which thus validates the performance-complexity trade-off discussed in Section \ref{sec_sub}. In practice, a suitable value of $\kappa$ needs to be determined based on the spatial channel correlation among neighboring grid points, where larger $\kappa$ is generally desirable when the correlation is high and thus the quantization loss is small. This is an interesting problem to be further studied in our future work.
\vspace{-1mm}
\section{Conclusions}
\vspace{-1mm}
This paper investigated the offline path planning for a cellular-connected UAV for minimizing its flying distance from given initial to final locations, subject to a communication quality constraint specified by the large-scale channel gain with its associated GBSs. We presented a new path optimization framework on the basis of the radio map, which characterizes the large-scale channel gains between each GBS and uniformly sampled locations on a 3D grid. Based on the radio maps, the optimal solution was obtained by solving an equivalent SPP, and a suboptimal solution with lower complexity was proposed based on a grid quantization method. Numerical results validated the efficacy of both proposed optimal and suboptimal solutions, and showed their performance-complexity trade-off.

\begin{spacing}{0.99}
\bibliographystyle{IEEEtran}
\bibliography{radiomap}
\end{spacing}
\end{document}